\documentclass{aa}
\usepackage{natbib}
\usepackage{graphics}
\bibpunct{(}{)}{;}{a}{}{,} 
\usepackage{times}

\newcommand{\Msun}{\ensuremath{M_{\sun}}}                         
\newcommand{\Rsun}{\ensuremath{R_{\sun}}}                         
\newcommand{\Teff}{\ensuremath{T_{\rm eff}}}                      
\newcommand{\FeH}{${\rm [Fe/H]}$} 
\newcommand{\MoH}{${\rm [M/H]}$}  
\newcommand{\kms}{\,km\,s$^{-1}$}                                 
\newcommand{\mc}[1]{\multicolumn{2}{c}{#1}}

\def\d{\displaystyle}
\def\kms{km\,s$^{-1}$}

\def\vs{$v\sin{i}$}
\def\te{$T_{\rm eff}$}
\def\lg{$\log{g}$}

\def\te{$T_{\rm eff}$}
\def\lgg{$\log{g}$}

\def\kic{\object{KIC 10661783}}
\def\cd{c\,d$^{-1}$}

\begin{document}

\title{Physical properties of the eclipsing $\delta$\,Sct star \kic\thanks{Based on observations made with the 
2-m Alfred Jensch telescope at the Th\"uringer Landessternwarte (TLS) Tautenburg and with the Mercator telescope,  
operated on La Palma by the Flemish Community at the Spanish Observatorio del Roque de los Muchachos of the 
Instituto de Astrof{\'{\i}}sica de Canarias.}}

\author{H.\ Lehmann\inst{1}  
\and J.\ Southworth\inst{2}  
\and A.\ Tkachenko\thanks{Postdoctoral Fellow of the Fund for Scientific Research (FWO), Flanders, Belgium}\inst{3} 
\and K.\ Pavlovski\inst{4}}

\institute{
Th\"uringer Landessternwarte Tautenburg, 07778 Tautenburg, Germany,
\email{lehm@tls-tautenburg.de} \and
Astrophysics Group, Keele University, Staffordshire, ST5 5BG, UK,
\email{astro.js@keele.ac.uk} \and
Instituut voor Sterrenkunde, KU Leuven, Celestijnenlaan 200D, B-3001 Leuven, Belgium,\\
\email{andrew@ster.kuleuven.be}  \and
Department of Physics, University of Zagreb, Bijenicka cesta 32, 10000 Zagreb, Croatia,
\email{pavlovski@phy.hr}}

\date{Received ; accepted}

\abstract
{\object{KIC 10661783} is an eclipsing binary that shows $\delta$\,Sct-like oscillations. More than 60 pulsation 
frequencies have been detected in its light curve as observed by the {{\it Kepler}} satellite.}
{We want to determine the fundamental stellar and system parameters of the eclipsing binary as a precondition 
for asteroseismic modelling of the pulsating component and to establish whether the star is a semi-detached 
Algol-type system.} 
{ We measured the radial velocities of both components from new high-resolution spectra using {\sc todcor} 
and compute the orbit using {\sc phoebe}. We used the {\sc korel} program to decompose the observed 
spectra into its components, and analysed the decomposed spectra to determine the atmospheric parameters. 
For this, we developed a new computer program for the normalisation of the {\sc korel} output spectra. 
Fundamental stellar parameters are determined by combining the spectroscopic results with those from the 
analysis of the {\it Kepler} light curve.}
{We obtain \te, \lg, \vs, and the absolute masses and radii of the components, together with their flux 
ratio and separation. Whereas the secondary star rotates synchronously with the orbital motion, the primary 
star rotates subsynchronously by a factor of 0.75. The newly determined mass ratio of 0.0911 is higher than 
previously thought and means a detached configuration is required to fit the light curve.}
{With its low orbital period and very low mass ratio, the system shows characteristics of the R\,CMa-type stars 
but differs from this group by being detached. Its current state is assumed to be that of a detached post-Algol 
binary system with a pulsating primary component.}

\keywords{Asteroseismology -- Stars: binaries: eclipsing -- Stars: variables: 
general -- Stars: atmospheres -- Stars: abundances}

\maketitle

\section{Introduction}

The {\it Kepler} satellite delivered light curves (LCs) of thousands of pulsating stars and opened, 
with its unprecedented photometric accuracy and long sequences of continuous observations, a new 
window for asteroseismology. An asteroseismic modelling analysis requires accurate fundamental 
parameters of the stars that cannot be gathered from the single-band photometry of {\it Kepler} alone. 
Ground-based, multi-colour, or spectroscopic follow-up observations can provide this missing information.

Eclipsing binaries (EBs) are outstanding targets for determining the fundamental parameters of stars from 
a combined LC and spectroscopic analysis. \kic\ is a short-period binary star that was found to be an 
EB by \citet{1997AcA....47..467P}. It was chosen by \citet{2011MNRAS.414.2413S} (hereafter Paper\,I) 
as a target for detecting of pulsations in the component stars of EBs, and {\it Kepler} observations 
of both long and short cadence were obtained. These observations were augmented with SuperWASP photometry 
in order to precisely measure the orbital period of the system. The residual LC after subtracting the 
eclipse variability was subjected to a frequency analysis, revealing at least 68 frequencies of which 
55 can be attributed to pulsation modes. The main frequency range lies between 18 and 31\,d$^{-1}$, and 
the variability due to pulsations could be assigned to the primary component of the system. 

In their modelling of the {\it Kepler} LC, the authors were able to define two possible solutions: 
that of a detached EB with a mass ratio of $q = 0.25$ and that of a semi-detached EB (secondary star 
filling its Roche lobe) with $q = 0.0626$. The latter option was preferred in light of preliminary 
spectroscopic measurements that favoured a low mass ratio. This implied that \kic\ was a member of 
the class of `oEA' stars \citep{2002ASPC..259...96M, 2003ASPC..292..113M}: active Algol-type systems 
where the primary component shows $\delta$\,Scuti-type pulsations. Moreover, it would be the oEA star 
with by far the richest pulsation spectrum detected until now. In the absence of radial velocity 
measurements, they were unable to measure the physical properties of the two stars.

The analysis of the {\it Kepler} LC of this short-period ($P = 1.2313622$\,d) EB, after removing 
the pulsation signatures, encountered some problems. A reasonable fit to the data could only be obtained 
by assuming a higher albedo for the primary than physically expected and an unusually low mass ratio of 
$q = 0.0626$. Together with the short period, these properties would make \kic\ a member of the subgroup 
of R\,CMa stars among the Algol-type stars (see \citealt{1999AJ....117.2980V} for a discussion, 
\citealt{2011MNRAS.418.1764B} for \object{R CMa}). The mass ratio is a critical piece of information when 
assessing wether the secondary 
component fills its critical Roche lobe (RL) or not, i.e.\ if the system is semi-detached or detached. 
It was derived in Paper\,I from the LC alone, and thus depended on the reliability of the model fit to 
the data.

In this paper we obtain and analyse new spectroscopic observations, which we use to improve the 
photometric model and to directly determine the absolute masses and radii of the components. We 
investigate the hypothesis that \kic\ is a semi-detached system, and determine the effective temperatures 
and chemical abundances of the stars from analysis of the separated spectra of the components.

\section{Observations}

\begin{table}
\tabcolsep 1.65mm
\caption{Number $N$ of spectra obtained in different nights in 2010.}
\begin{tabular}{|cccc|cccccc|}
\hline
\multicolumn{4}{|c|}{TLS} & \multicolumn{6}{c|}{HERMES}\\
date & $N$ & date & $N$ & date & $N$ & date & $N$ & date & $N$\\
\hline
07/10 & 5 & 07/16 & 4 & 07/22 & 6 & 07/26 & 5 & 07/30 & 5\\
07/11 & 2 & 07/18 & 4 & 07/23 & 5 & 07/27 & 5 & 07/31 & 6\\
07/13 & 3 & 07/19 & 6 & 07/24 & 5 & 07/28 & 5 & 08/01 & 6\\
07/15 & 2 &       &   & 07/25 & 6 & 07/29 & 5 &       &  \\
\hline
\end{tabular}
\label{t1}
\end{table}

In 2010, we obtained 26 high-resolution spectra of KIC 10661783 in seven almost consecutive nights using the 
Coud\'e-\'echelle spectrograph at the 2-m telescope of the Th\"uringer Landessternwarte (TLS) Tautenburg, 
Germany, and 59 spectra in 11 consecutive nights with the HERMES spectrograph \citep{2011A&A...526A..69R} 
at the 1.25-m Mercator telescope at La Palma, Spain. Table\,\ref{t1} gives the journal of observations.

The TLS spectra have a resolving power of 64\,000 and cover the wavelength range 4700--7400\,\AA. The 
exposure time was 30\,min and the spectra have a signal-to-noise ratio (SN) of 90 on average. The HERMES 
spectra have a resolving power of 85\,000 and cover 3740--9000\,\AA. The exposure time was 40 \,min and the mean 
SN was 130. The higher SN obtained with the HERMES spectrograph is due to its higher efficiency and 
the better weather and seeing conditions at La Palma.
 
The TLS spectra were reduced using standard {\sc midas} packages and a routine for the calibration of the 
instrumental radial velocity (RV) zero-point using O$_2$ telluric lines. The idea of using telluric 
absorption lines for obtaining accurate RV measurements was proposed by \cite{1973MNRAS.162..243G}. 
\cite{2010A&A...515A.106F,2012MNRAS.420.2874F} showed, based on a study done for the HARPS spectrograph at 
La Silla, that O$_2$ telluric lines are stable down to 10\,m\,s$^{-1}$ on a timescale of 6 years and that 
atmospheric phenomena introduce variations at the 1--10\,m\,s$^{-1}$ level (rms). One complication is that 
the telluric bands cover only certain regions of the observed spectra. From our experience with the TLS 
spectrograph over several years, we can reach an accuracy of better than 50\,m\,s$^{-1}$ in this way. For 
the reduction of the HERMES spectra we used the HERMES reduction pipeline, followed by the normalisation of 
the spectra to the local continuum based on a separate routine.

\section{Investigation of the primary component}

Throughout the paper, we designate the brighter component of \kic\ (the former mass-gainer according to the 
standard evolutionary scenario for Algol systems) as the primary star and the low-mass companion (the donor) 
as the secondary star.

\subsection{Radial velocities and atmospheric parameters\label{AtmoPar}}

In this first approach, we ignored the faint contribution of the secondary component to the composite spectra. 
RVs were measured from the composite spectra by cross-correlation in two steps. In the first step, we 
determined the RVs to build a mean spectrum from the observed ones. We started with one arbitrary spectrum of 
 reasonably high SN  as the template and determined the RVs on a relative scale. Then we shifted all the observed spectra 
according to the measured RVs, co-added them and repeated the RV measurement using the averaged, mean spectrum 
as the template. After an iteration, we fitted the averaged spectrum by synthetic spectra calculated on a grid 
of atmospheric parameters and determined the RVs from cross-correlation of the observed spectra with the 
best-fitting synthetic spectrum on an absolute scale. The synthetic spectra were calculated with {\sc SynthV} 
\citep{1996ASPC..108..198T} based on model atmospheres computed with {\sc LLmodels} \citep{2004A&A...428..993S}. 
The line tables were taken from the VALD data base \citep{2000BaltA...9..590K}. The fitting procedure was 
described in detail in \citet{2011A&A...526A.124L}. Finally, we repeated the procedure with the new, 
synthetic template, iteratively improving the RVs as well as the mean spectrum of \kic.

\begin{table}
\tabcolsep 1.7mm
\caption{Atmospheric parameters derived from the mean composite spectrum of \kic.}
\begin{tabular}{ccccc}
\hline\hline\vspace{1mm}
\te & \lg & [M/H] & \vs & $\zeta_{\rm turb}$\\
(K)  &      &       & (\kms) & (\kms)\\
\hline
7\,664$\pm$49 & 3.45$\pm$0.18 & $-$0.36$\pm$0.06 & 80.6$\pm$0.7 & 2.9$\pm$0.2\\
\hline
\end{tabular}
\label{t2}
\end{table}

From the final mean spectrum, we determined the atmospheric parameters listed in Table\,\ref{t2}. Although
these results have been derived from the composite spectrum of the binary, they are good first estimates of the 
stellar parameters of the primary star because the secondary star contributes only a small fraction of the light 
in the spectrum (see Sect.\,\ref{Both}).

\subsection{Orbital solution}

We used the method of differential corrections to the orbital elements \citep{1910PAllO...1...33S,1941PNAS...27..175S} 
to determine the orbit of the primary component from the observed composite spectra. After deriving the solution from the complete 
data set, we fixed the obtained parameters and calculated the O-C residuals for the TLS and HERMES RVs separately. 
The difference between the two data sets is 1.615\,\kms\ and we added this value to the HERMES RVs to calculate the 
final orbital solutions. Including all the RVs and allowing for an eccentric orbit, we obtain an eccentricity of 
$e = 0.036 \pm 0.001$ from all RVs and $0.016 \pm 0.006$ if we omit the RVs around primary eclipse. No significant 
non-circularity could be found after also rejecting all spectra near to secondary eclipse. 
From now on we assume a circular 
orbit for this close binary system and fix the orbital period to the value of 1.23136220\,d obtained in Paper\,I. 
This value is two orders of magnitude more precise than that which can be determined from our spectroscopic 
observations. Fig.\,1 shows the RVs folded with the orbital period. In the O-C residuals (Fig.\,\ref{f1a}) we 
clearly see the Rossiter-McLaughlin effect \citep{1924ApJ....60...15R,1924ApJ....60...22M} around Min\,I. This 
will be discussed in Sect.\,\ref{TODCOR}, based on the analysis of the individual components. The resulting orbital 
parameters are listed in Table\,\ref{t3} given in Sect.\,\ref{Comp} and will be discussed there. 
The quoted errorbars are formal 1$\sigma$ limits computed from the inverse matrix of the least squares problem of 
the differential-correction approach.

We used the {\sc period04} program \citep{2004IAUS..224..786L}) to search the O-C residuals of the spectroscopic 
orbital solution for periodic variations. No significant contributions could be found up to the Nyquist frequency of 
8.37\,\cd.

\begin{figure}
\includegraphics[angle=-90, width=9cm]{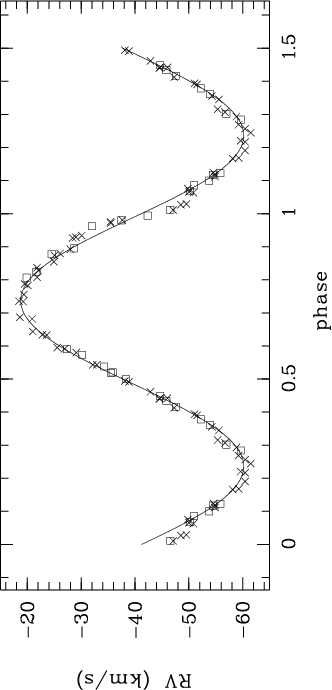}
\caption{RVs of the primary component measured from TLS (squares) and HERMES (crosses)
spectra folded with the orbital period. The solid curve is based on the
parameters listed in Table\,\ref{t3}. Phase zero corresponds to Min\,I.}
\label{f1}
\end{figure}

\begin{figure}\vspace{-4mm}
\includegraphics[angle=-90, width=9cm]{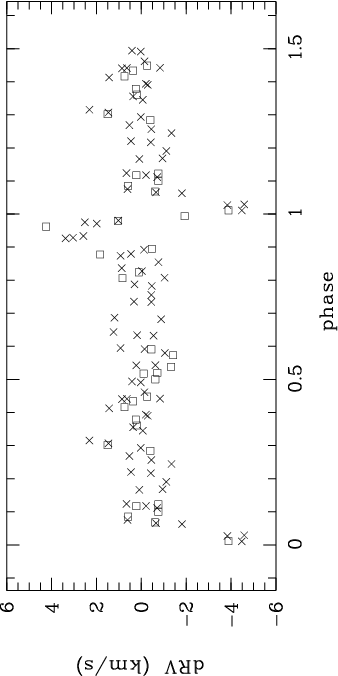}
\caption{The O-C residuals of the fit in Fig.\,\ref{f1}, indicating the presence of the Rossiter-McLaughlin effect 
during primary eclipse.}
\label{f1a}
\end{figure}

\section{Investigation of both components\label{Both}}

\subsection{The impact of the secondary star}

For a visual search for the lines of the secondary, we arranged all spectra in a two-dimensional frame (Fig.\,\ref{f2}), 
averaging them into 30 orbital phase bins. The impact of the secondary star could be only detected in a few iron 
lines. It is most prominent (although still very faint) in the \ion{Fe}{ii} 4957\,\AA\ line. Fig.\,\ref{f2} shows 
this line where the central line depth is about 13\% for the primary and about 2\% for the secondary. The lines of the 
secondary could only be found in the HERMES spectra, which have a higher SN than the TLS spectra.

\subsection{Decomposing the spectra with {\sc korel}}

We used the Fourier-based {\sc korel} program \citep{1995A&AS..114..393H,2006A&A...448.1149H} to decompose the 
spectrum of \kic. It delivers the decomposed spectra normalised to the common continuum of the composite 
input spectrum, together with the optimised orbital elements. In the case 
of light variability like in EBs, one can also calculate variable line strengths (mean over all included lines) for 
both components separately. 

\begin{figure} \centering
\includegraphics[angle=-90, width=6.33cm]{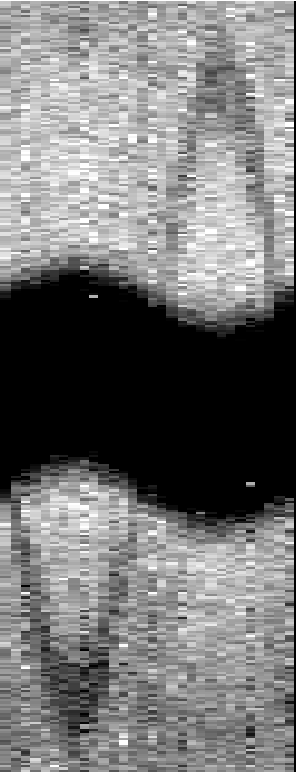}
\caption{The strong and faint \ion{Fe}{ii} 4957\,\AA\ lines of the primary and secondary,
respectively. The vertical axis gives the orbital phase of observation, and secondary 
eclipse occurs at the centre of the image. The elongation in RV (horizontal axis) 
corresponds to the velocity semiamplitudes given in Table\,\ref{t3}.}
\label{f2}
\end{figure}

\begin{figure*}
\includegraphics[width=18.3cm]{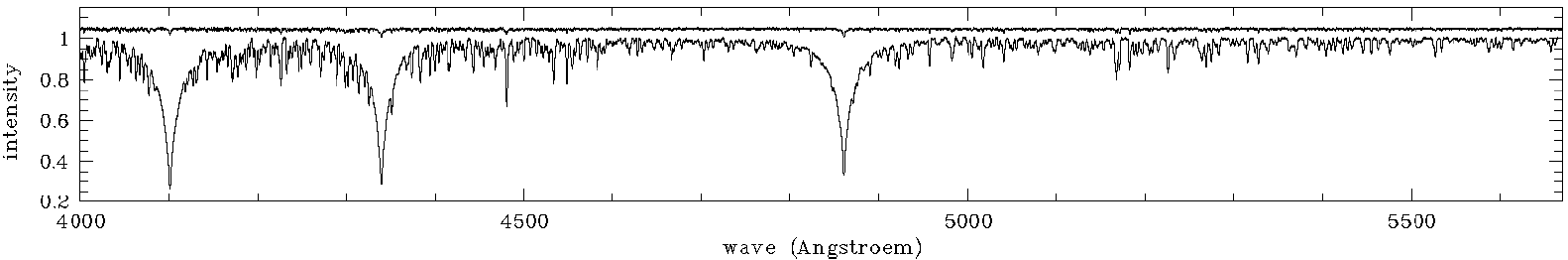}
\caption{The decomposed spectra of the primary and secondary components. 
The latter is shifted upwards by 0.05 for clarity.}
\label{BothSpectra}
\end{figure*}

\begin{table*}
\tabcolsep 2.2mm
\caption{Elemental abundances of the primary ($A_1$) and the secondary ($A_2$) compared to the 
Sun and their 1$\sigma$ errors. Solar values are given below the element designations.}
\begin{tabular}{lrrrrrrrrrrrrr}
\hline \hline
      & \multicolumn{1}{c}{C}    
      & \multicolumn{1}{c}{N}     
      & \multicolumn{1}{c}{O} 
      & \multicolumn{1}{c}{Mg}     
      & \multicolumn{1}{c}{Si}    
      & \multicolumn{1}{c}{Ca}   
      & \multicolumn{1}{c}{Sc}    
      & \multicolumn{1}{c}{Ti}    
      & \multicolumn{1}{c}{V}    
      & \multicolumn{1}{c}{Cr}    
      & \multicolumn{1}{c}{Mn}    
      & \multicolumn{1}{c}{Fe}    
      & \multicolumn{1}{c}{Ni}\\
      & $-3.65$ &  $-4.26$ &  $-3.38$ &  $-4.51$ &  $-4.53$ &  $-5.73$ &  $-8.99$ &  $-7.14$ &  $-8.04$ &  $-6.40$ &  $-6.65$ &  $-4.59$ &  $-5.81$\\
\hline
$A_1$             &$-0.14$&$+1.21$&$-0.02$&$+0.10$&$-0.11$&$-0.19$&$+0.34$&$+0.05$&$+0.38$&$+0.11$&  --   &$-0.04$&$-0.33$\\
$\sigma_{\rm A1}$ &0.28   &  0.34 &  0.50  &  0.22 &  0.44 &  0.25 &  0.43 &  0.14  &  0.33 &  0.17 &  --   &  0.07  &  0.33 \\
$A_2$             &$-0.79$&  --   &  --    &$-0.35$&  --   &$-0.43$&  --   &$-0.21$& --    &$-0.28$&$-0.15$&$-0.34$&$-0.42$\\
$\sigma_{\rm A2}$ & 0.25  &  --   &  --    &  0.29 &  --   &  0.47 &  --   &  0.18  & --    &  0.24 &  0.44 &  0.10  &  0.28 \\
\hline
\end{tabular}
\label{t5}
\end{table*}

One problem in applying the {\sc korel} program is that the zero frequency mode in the Fourier domain is unstable 
and cannot be determined from the Doppler shifts alone. This effect gives rise to low-frequency undulations in the 
continuum of the decomposed spectra and can prevent an accurate determination of the local continuum. We therefore 
followed the procedure discussed by \citet{2008A&A...482.1031H}, which shows that inherent light variations can 
stabilise the separation of the stellar components in the {\sc korel} method, and allowed for variable line strengths 
in the {\sc korel} solutions. In this way, the degeneracy in the determination of the local continuum should be lifted 
by the variation of the light ratio. However, we were surprised by our first analysis, that resulted in two identical 
decomposed spectra. The calculated line strengths of both components were extraordinarily strong at some single epochs 
close to the eclipses and very weak over all other phases. We assume that most of the spectra were included with zero 
weight into the calculations. One reason that the variable weighting of spectra failed could be, besides the
faintness of the second component, the strong 
Rossiter-McLaughlin effect during primary eclipse.  It deforms the lines of the primary 
during the eclipse (see Sect.\,\ref{TODCOR}) which is not taken into account by {\sc korel}.

In a second attempt, excluding the eclipse phases and using constant line strengths, we ended up with a reliable 
solution. We used the HERMES spectra only, starting with the iron line shown in Fig.\,\ref{f2}, and expanded the 
wavelength range we used up to a limit where we obtained the smallest scatter of the RVs of both components. From 
this region, 4900--5193\,\AA, we calculated the orbital elements listed in Table\,\ref{t3} and used them as fixed 
parameters for the spectral disentangling that we performed on a wider spectral range. Because the continuum calculated 
during the normalisation of the observed spectra is very uncertain at the blue edge of the spectra, we restricted the 
wavelength range for the spectral disentangling to $\lambda > 4000$\,\AA. 

As noted above, {\sc korel} introduces undulations in the calculated continuum of the decomposed spectra. To 
overcome this problem, we split the spectra into overlapping wavelength bins of 50\,\AA\ width, with wider bins 
for the Balmer lines, and corrected the continuum of each of the resulting decomposed parts using spline functions. 
The final decomposed and corrected spectra were then built by merging all the short, decomposed regions and averaging 
the overlapping parts. It can be seen from Fig.\,\ref{BothSpectra} that the resulting spectra do not suffer 
from any continuum undulations.

\begin{figure}\centering
\includegraphics[angle=-90, width=7.5cm]{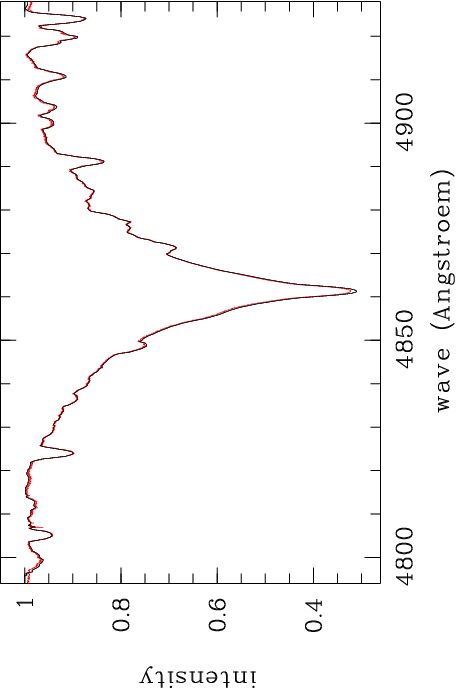}
\caption{Comparison of the decomposed H$\beta$ line profiles of the primary 
component obtained with {\sc cres} (black) and with {\sc korel} (red).}
\label{Hbeta}
\end{figure}

One remaining problem is the low accuracy of the calculated continuum in the region of the broad Balmer lines of the 
primary, where we had to use larger wavelength regions for the disentangling. The problem is not as serious for the 
secondary component because the Balmer lines of the cooler star are much narrower. To check for the accuracy achieved, 
the {\sc cres} program\footnote{http://sail.zpf.fer.hr/cres/} \citep{2004ASPC..318..107I}  was applied to decompose 
the H$\beta$ and H$\gamma$ line profiles. {\sc cres} uses singular value decomposition in the wavelength domain to 
disentangle spectra. The agreement of the resulting profiles with those obtained with the {\sc korel} program was 
excellent. This is illustrated by Fig.\,\ref{Hbeta} which shows a comparison of the H$\beta$ line profiles of the 
primary component obtained with both methods where the {\sc cres} decomposed spectrum looks less noisy because it 
was rebinned to a smaller resolution to save computing time. 

\subsection{Analysis of the decomposed spectra\label{SpecAna}}

For analysis of the decomposed spectra, we used the same approach as in Sect.\,\ref{AtmoPar} but extended our program 
by including the computation of the wavelength-dependent flux ratio between the components, which is needed to 
normalise the {\sc korel} output spectra. A similar and more sophisticated method was described in 
\citet{2005A&A...439..309P}. We used the approach introduced by \citet{2012IAUS..282..395L}, which is described in 
detail in Appendix\,A. For the wavelength dependence of the continuum flux ratio $f=F_2/(F_1+F_2)$, we used the linear 
approach:
\begin{equation}
f = f_0+f_1\frac{\lambda-\lambda_0}{\d\lambda_0}
\label{fluxeq}
\end{equation}

\begin{figure}\centering
\includegraphics[angle=-90, width=8cm]{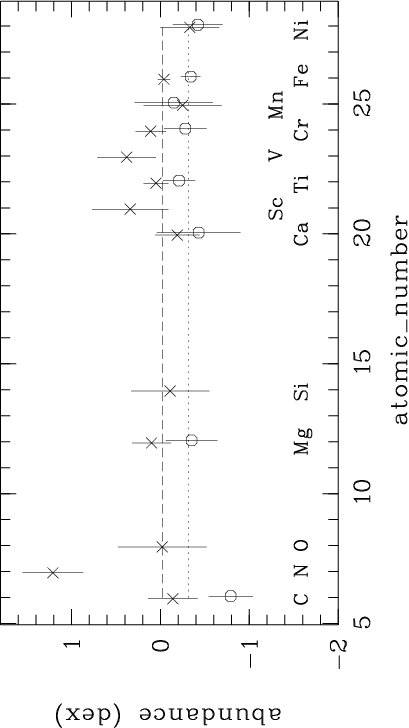}
\caption{Chemical abundances of the primary (crosses) and the secondary (open circles) components, 
relative to solar. Chemical elements are sorted by their atomic numbers. 
The dashed and dotted lines show the weighted mean for the primary and secondary stars, respectively.}
\label{Abund_figure}
\end{figure}

In a first step, we determined \te, \MoH, and \vs\ together with the flux ratio and continuum corrections. Because 
of the observed ambiguity between \lg\ and the other parameters we fixed \lg\ of the components to the values derived 
from the combined LC and RV analysis after iteration (see Table\,\ref{tab:wdf} in Sect.\,\ref{Combined}). Then we 
renormalised the decomposed spectra according to the derived flux ratio, applying the continuum corrections as described 
in Appendix\,A, and analysed the renormalised spectra separately. Starting with the abundances of all elements scaled 
to the derived \MoH, we determined the optimised abundances of each element step by step, including only those elements 
where significant contributions could be found in the wavelength range under consideration. We did not try to adjust 
the helium abundance because we do not observe any He lines in this region. For the cooler secondary star, we included 
molecules in the spectral line tables and were therefore able to determine its carbon abundance by fitting the CH bands. 

Table\,\ref{t5} lists the determined abundances. The uncertainties were calculated from the hypersurface of the 
$\chi^2$ values of all grid points in \te, \vs, and the abundance of the corresponding element 
\citep[see][]{2011A&A...526A.124L} and include possible correlations between these parameters.

Fig.\,\ref{Abund_figure} shows the abundances with their $1\sigma$ error bars. The dashed and dotted lines show 
the weighted means calculated from all elements except for N for the primary and C for the secondary. It can be seen 
that the abundances of all other elements agree with the weighted mean to within $1\sigma$. We conclude that N is 
strongly overabundant in the primary and C is underabundant in the secondary star. 

Table\,\ref{t6} lists the derived atmospheric parameters where \FeH\ is taken from Table\,\ref{t5} and \MoH\ 
represents the mentioned weighted mean. 
In the lower part of Table\,\ref{t6}, we list the coefficients of the flux ratio function as defined in 
Eq.\,\ref{fluxeq}.

\begin{table}\centering
\caption{Atmospheric parameters from spectrum analysis.}
\begin{tabular}{llcc}
\hline\hline
                  &           & primary         & secondary       \\
\hline
\te               & (K)       & $7764 \pm 54$   & $5980 \pm 72$   \\     
\lgg              & (cgs)     &  3.9 fixed      &  3.6 fixed      \\  
\vs               & (\kms)    & $79 \pm 4$      &   $48 \pm 3$    \\
\FeH              & (dex)     & $-0.04\pm0.07$  & $-0.34\pm0.10$  \\
\MoH              & (dex)     & $-0.02\pm0.05$  & $-0.31\pm0.08$  \\
\hline
\multicolumn{2}{c}{$f_0(\lambda_0$=5000\,\AA)} & \multicolumn{2}{c}{0.062564} \\
\multicolumn{2}{c}{$f_1(\lambda_0$=5000\,\AA)} & \multicolumn{2}{c}{0.159515} \\
\hline
\end{tabular}
\label{t6}
\end{table}

\subsection{Orbital solution with {\sc todcor}\label{TODCOR}}

The {\sc korel} code  determines the orbit directly from the observed composite spectra. The RVs 
calculated with {\sc korel} are not independent measurements, but  represent the shifts applied 
to the spectra.  They are optimised together with the orbital elements assuming Keplerian motion. 
An additional aspect of the {\sc korel} solution is that the spectra close 
to primary minimum had to be excluded from the calculations. We therefore used the {\sc todcor} program 
\citep{1992ASPC...32..164M,1994Ap&SS.212..349M,1994ApJ...420..806Z} together with the HERMES spectra of \kic\ 
to obtain independent measurements of the RVs of both components. 
 
\begin{table*}\centering
\tabcolsep 3mm
\caption{Orbital parameters of the binary system (see text).}
\begin{tabular}{lcccccc}
\hline\hline
       & $\gamma$ (\kms)  & $K_1$ (\kms)     & $K_2$ (\kms)     & $q$                  & $a$ (\Rsun)  & $i$ ($^\circ$) \\
\hline       
CCF    & $-39.45\pm 0.80$ & $20.66\pm 0.18$   & --               & --                   & --               & --            \\
{\sc korel}  & --               & $21.44\pm 0.21$   & $235.9\pm 5.3$   & ($0.0909\pm 0.0011$)   & ($6.31\pm 0.14$) & --            \\ 
{\sc todcor} & $-41.04\pm 0.10$ & ($21.67\pm 0.18$) & ($237.9\pm 1.1$) & $0.09109\pm 0.00065$ & $6.370\pm 0.026$ & $82.3\pm 1.3$ \\
WD2004 & --               & --                & --                & 0.09109 fixed & $6.375\pm 0.027$ & $82.39\pm 0.23$ \\
\hline
\end{tabular}
\label{t3}
\end{table*}
 
We applied an implementation of {\sc todcor} that is able to determine the RVs from composite spectra together 
with the flux ratio between the components. We used two synthetic template spectra, calculated for the \te, \lg, 
and \vs\ listed in Table\,\ref{t6}. The resulting RVs of the primary were unsatisfactory, because we could 
not find any sign of the Rossiter-McLaughlin effect (RME), despite this being clearly discernable in the RVs that 
we obtained from cross-correlation (see Fig.\,\ref{f1a}). 

Considering the RME, we must be aware that we are dealing with two different types of velocity. One type is the 
classical RV, defined as the velocity component of the centre of mass of the star along the line of sight. For 
binary stars this is due to Keplerian motion, and can easily be measured from cross-correlation or line profile 
fitting when the lines are symmetrically broadened by rotation. The symmetry is lifted during the eclipses and the 
observed spectrum results from the line profiles emerging from single surface points, each RV-shifted by rotation 
according to its position on the stellar disc, integrated over the visible, non-eclipsed part of this disc. What 
we call RV in this case strongly depends on how the positions of the resulting asymmetric lines are measured. 
 Next, we wanted to estimate the expected amplitude of the RME for \kic, and the degree of asymmetry 
of the line profiles caused by the RME.

\begin{figure}
\includegraphics[angle=-90, width=4.35cm]{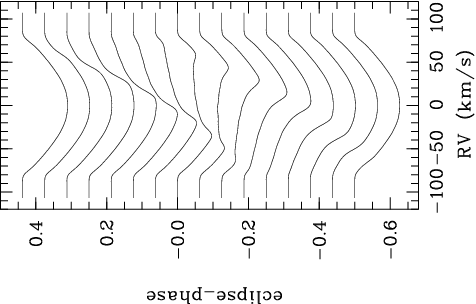}
\includegraphics[angle=-90, width=4.35cm]{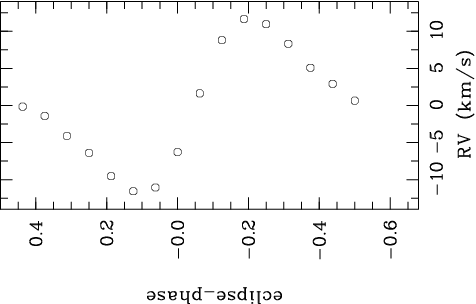}
\caption{The Rossiter-McLaughlin effect: Synthetic line profiles computed for different phases around primary eclipse
(left) and the RVs computed from these profiles (right).}
\label{Rossiterprofiles}
\end{figure}

\cite{2005ApJ...622.1118O} developed analytical formulae to estimate the amplitude of the RME for transiting planets. 
These formulae are only valid for the case that the radius of the planet is much smaller than that of its host star. 
 Instead, we used a simple model consisting of a rotating star eclipsed by a dark 
disc moving across the disc of the star at a latitude corresponding to the derived orbital inclination and separation 
of \kic\ (see Table\,\ref{tab:wdf}). We divided the stellar disc into 100 by 100 surface elements. For each of these 
elements we computed the emerging line profile assuming Gaussians for the intrinsic profiles, shifted in RV by 
rotation. The intensity was scaled according to a linear limb darkening law. All parameters (separation, inclination, 
radii, and \vs) were adjusted to our final measurements of these values (Tables \ref{t3} and \ref{tab:wdf}). 
Fig.\,\ref{Rossiterprofiles} shows the profiles obtained from an integration across the visible, uneclipsed disc. 
We define the eclipse phase to be $-0.5$ at the beginning and $+0.5$ at the end of the eclipse. It can be seen that 
the impact of the RME on the line profiles causes a shift of the line centre (or centre of gravity) in wavelength 
(or RV) which is due to the varying shape of the profiles. The lines get highly asymmetric during ingress and egress 
and the star shows a double-peaked profile at the centre of the (partial) eclipse. The RVs shown in the right panel 
of Fig.\,\ref{Rossiterprofiles} are computed from the centres of gravity of the profiles. Thus, we expect an 
amplitude in RV of the order of $\pm 10$\,\kms\ and a varying effective line width having minima during ingress and 
egress of primary eclipse and a maximum at its centre.

\begin{figure}
\includegraphics[angle=-90, width=9cm]{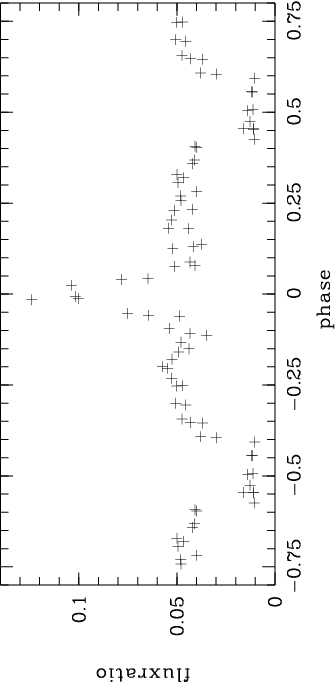}\vspace{-7mm}
\includegraphics[angle=-90, width=9cm]{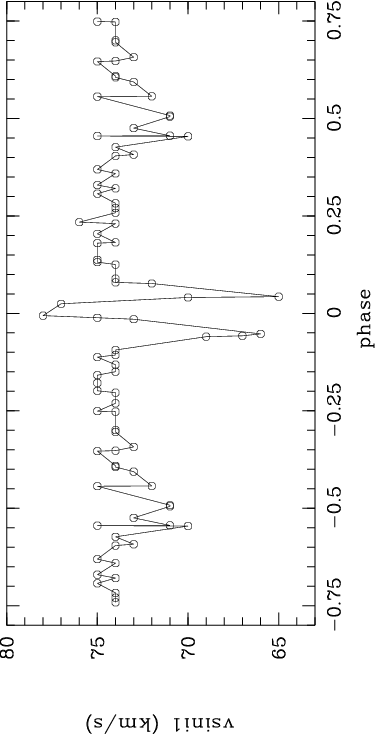}
\caption{Flux ratio (top) and \vs\ of the primary component (bottom) determined with {\sc todcor}, folded with the 
orbital period.
Phase zero corresponds to Min\,I.}
\label{todcor_flux}
\end{figure}

\begin{figure}
\includegraphics[angle=-90, width=9cm]{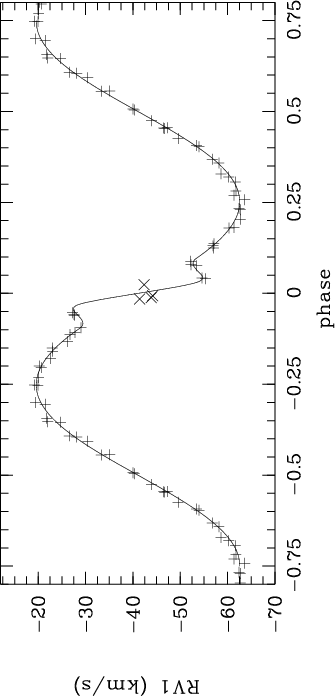}\vspace{-7mm}
\includegraphics[angle=-90, width=9cm]{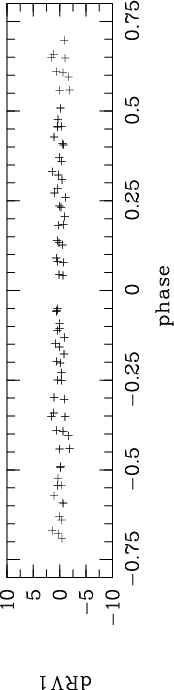}
\caption{RVs of the primary component determined with {\sc todcor}.
The solid curve shows the orbital solution calculated with {\sc phoebe}. 
RVs included into the solution are shown by plus signs, outliers by crosses.
The lower panel shows the O--C residuals.}
\label{todcor_orbit1}
\end{figure}

\begin{figure}
\includegraphics[angle=-90, width=9cm]{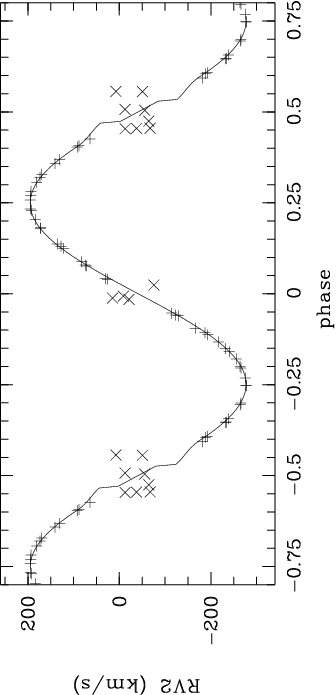}\vspace{-7mm}
\includegraphics[angle=-90, width=9cm]{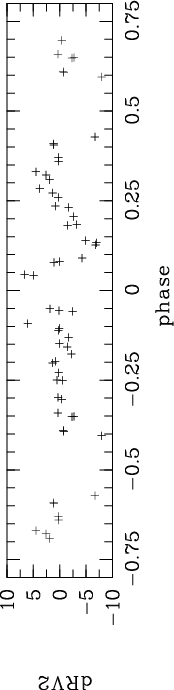}
\caption{As Fig.\,\ref{todcor_orbit1} but for the secondary component.}
\label{todcor_orbit2}
\end{figure}

 To account for the latter effect, we computed the RVs with {\sc todcor} in a second step 
 from each spectrum by using different \vs\ values for each component and optimising the \vs\ by maximising
the correlation coefficient. The results 
are shown in Figures \ref{todcor_flux} to \ref{todcor_orbit2}. 
We used a grid of synthetic spectra with a 1\,\kms\ step width in \vs. The optimum values for the primary component are shown 
in the lower panel of Fig.\,\,\ref{todcor_flux}. Outside the primary eclipse we see a small scatter that increases 
around secondary minimum. The behaviour during primary minimum is exactly as expected and described above. The 
behaviour during secondary eclipse appears to be an artefact, due to line-blending effects between the two components 
and the fact that {\sc todcor} searches for a second component even when the eclipse is total. For the computation of 
the RVs we have set the \vs\ of the primary star outside primary eclipse to the mean value of 74.5\,\kms. This value 
is slightly lower than the value of $79\pm 4$\,\kms\ obtained from the analysis of the decomposed spectra 
(Sect.\,\ref{SpecAna}). The \vs\ values of the secondary component obtained in this way show pure scatter of about 
$\pm 2$\,\kms. For the final calculations, we fixed them to the mean value of 38\,\kms\ that is distinctly lower 
than the value of $48\pm 3$\,\kms\ following from the spectral analysis. A discussion  of this
result is given in the next two sections.

Figure\,\ref{todcor_flux} shows, in its upper part, the 
flux ratio obtained versus orbital phase. Its value outside the eclipses is slightly lower than the value of 0.063 
obtained from spectrum analysis (Sect.\,\ref{SpecAna}). It can be seen that the program recognises the increase of 
the flux ratio during primary eclipse and a decrease during secondary eclipse. For the computation of the RVs, we 
set the flux ratio for the central part of the secondary eclipse to zero since it is a total eclipse.

Figures\ \ref{todcor_orbit1} and \ref{todcor_orbit2} show the  obtained RVs folded on the orbital period. Values very 
close to the systemic RV could not be determined accurately, due to blending effects. Such values occur at the centre of the 
primary eclipse for the RVs of the primary and during both eclipses for the secondary and have been rejected from the 
computation of the orbit. The RME during primary eclipse can clearly be seen, is of the right magnitude compared to 
Fig.\,\ref{Rossiterprofiles}, and is much more pronounced than in Fig.\,\ref{f1} where we neglected the impact of 
the secondary component on the spectra.

 Finally, we used the {\sc phoebe} program \citep{2005ApJ...628..426P} for computing the orbit based on the RVs derived with 
{\sc todcor}. This program is able to include the RVs of both components into a common solution and also considers 
the RME and proximity effects. We used different weights for the RVs of the primary and secondary component. They 
were obtained from the final standard deviation of the corresponding O-C values, which are 0.75\,\kms\ for the RVs 
of the primary and 3.13\,\kms\ for those of the secondary. The orbital curves computed with {\sc phoebe} are shown 
in Figures\,\ref{todcor_orbit1} and \ref{todcor_orbit2}. Table\,\ref{t3} lists the results, together with the 
results from our previously described methods and the combined solution described in Sect.\,5.  
It lists the systemic velocity $\gamma$, the RV semiamplitudes of the components $K_{1,2}$, the mass ratio 
$q$, the orbital separation $a$, and the inclination $i$ derived with different methods. 
Orbital period and time 
of primary mid-minimum were fixed to the photometric values of 1.2313622\,d and 245\,506\,5.77701, respectively. 
Values that were not directly measured but were derived are given in parentheses.
The fitting of the
RME with {\sc phoebe} also allowed us to compute, based on the measured \vs, the values of orbital separation and inclination. The
result is in good agreement with the orbital solution obtained from the light curve analysis.

\subsection{Comparison of the different solutions\label{Comp}}

The RVs determined via cross-correlation can only give the RV semi-amplitude, $K_1$, of the primary component. 
Since we neglected the impact of the secondary on the composite spectra, the derived $K_1$ will slightly differ from 
the true value. Thus, the results have been only used to obtain the mean composite spectrum for a first analysis.

The orbital solution determined with {\sc korel} was derived by fitting the orbital parameters simultaneously 
with the optimum shifts (formal RVs). {\sc todcor} also uses the composite spectra but determines the RVs 
of the single stars by cross-correlating the observed spectra with two different synthetic templates in a 
two-dimensional plane. The orbital elements derived with {\sc todcor} show the smallest uncertainties. Whereas the 
uncertainties in $K_1$ obtained from {\sc korel} and {\sc todcor} are of the same order, the {\sc korel} 
uncertainty in $K_2$ is an order of magnitude higher. We assume that the low accuracy of $K_2$ derived with 
{\sc korel} also influences the determination of $K_1$ and adopt the orbital parameters that are based on the 
{\sc todcor} RVs as the final ones.

\begin{figure}
\includegraphics[angle=-90, width=9cm]{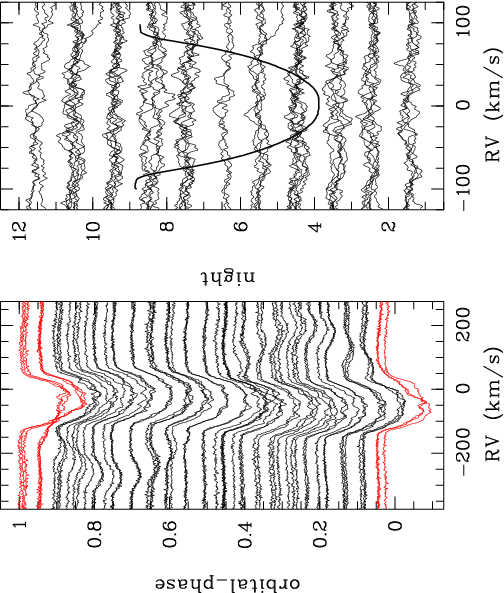}
\caption{Left: LSD profiles computed from all HERMES spectra, vertically arranged according to orbital phase. 
Profiles distorted by the RME are shown in red. Right: Differential LSD spectra from different nights (see text). 
The synthetic LSD profile of the primary is overplotted for a better distinction between line profile and continuum 
regions.}
\label{LSD_double}
\end{figure} 

We used {\sc todcor} mainly to derive the RVs of the components. Using optimised line widths, {\sc todcor} 
only fails to compute correct RVs for the primary star at the centre of primary minimum, where line blending effects
between both components occur and the spectral lines of the primary component 
are deformed by the RME. Although we notice that the line widths and flux ratio during primary eclipse are adequate, 
we do not trust the absolute values delivered by {\sc todcor}. Firstly, the flux ratio during primary eclipse is 
much too high compared to the eclipse depth (and should be zero during secondary eclipse). From the light curve 
(Fig.\,\ref{LC}) we expect an increase of the flux ratio by a factor of about 1.26, whereas {\sc todcor} 
(Fig.\,\ref{todcor_flux}) yields more than twice the flux ratio outside the eclipses. Secondly, the \vs\ derived 
from spectral analysis is more reliable because it exactly corresponds to the case of synchronised rotation.

\begin{figure}\centering
\includegraphics[angle=-90, width=7.5cm]{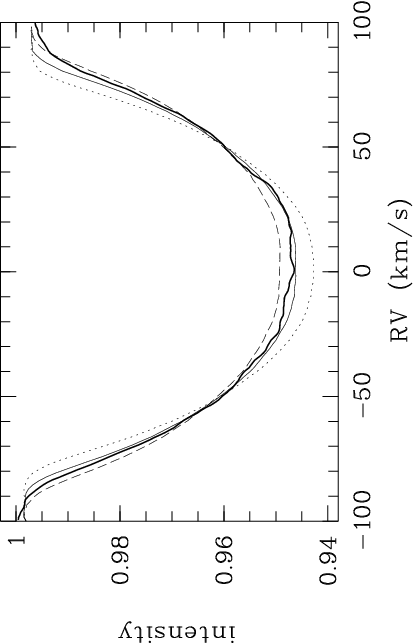}
\caption{Comparison of the LSD profile observed for the primary component (thick) with 
synthetic LSD profiles computed for \vs\ values of 75 (dotted), 80 (solid), and 85\,\kms\ (dashed).}
\label{LSD_comp1}
\end{figure}  
  
\begin{figure}\centering
\includegraphics[angle=-90, width=7.5cm]{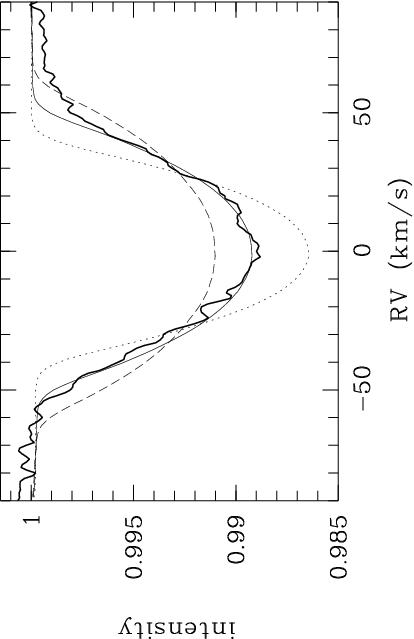}
\caption{Comparison of the LSD profile observed for the secondary component (thick) with 
synthetic LSD profiles computed for \vs\ values of 38 (dotted), 48 (solid), and 58\,\kms\ (dashed).}
\label{LSD_comp2}
\end{figure}

\subsection{LSD profiles}

To prove that the spectral disentangling delivered correct results and the \vs\ obtained from spectrum 
analysis give the accurate values, we analysed the observed instead 
of the decomposed spectra. To enhance the SN, we applied the least squares deconvolution (LSD) technique 
\citep{1997MNRAS.291..658D,1997MNRAS.291....1D} to each of the observed spectra in the range 4910--5670\,\AA\ 
(a region redwards of H$\beta$ with many metallic lines). LSD computes high-precision average line profiles from all 
lines in a spectrum, as shown in the left panel of Fig.\,\ref{LSD_double}. We selected 23 spectra taken at orbital 
phases of largest RV separation. The computed LSD profiles were shifted according to the orbital RVs of the primary 
and co-added to get the mean LSD profile of the primary component. The secondary was treated analogously. 
Figures\,\ref{LSD_comp1} and \ref{LSD_comp2} show the comparison of the observed mean profiles with synthetic LSD 
profiles computed from our atmospheric model (Table\,\ref{t2}) for different \vs. It can be seen that the \vs\ 
of 79\,\kms\ of the primary and 48\,\kms\ of the secondary derived from the analysis of the decomposed spectra 
fit the observed profiles very well. In particular, we can rule out the \vs\ of 38\,\kms\ corresponding to the mean 
value from {\sc todcor}. 

The right panel of Fig.\,\ref{LSD_double} shows the differential spectra after subtracting the best-fitting synthetic 
LSD profile from the observed ones (profiles observed during primary minimum are removed). 
The vertical axis gives the number of the observing night in a running order. It can be seen that the 
scatter in the range of the line profiles is distinctly larger than in the continuum regions and it can be assumed 
that the differential spectra show intrinsic signals from high-degree $l$ modes of non-radial pulsations. The 
investigation of this effect is beyond the scope of the current paper.

\section{Combined photometric-spectroscopic analysis\label{Combined}}

In Paper\,I we found two good solutions to the LC, one with a detached configuration and a mass ratio of $q = 0.25$ 
and one with a semi-detached nature and $q = 0.0626$. We rejected the former on the basis of the spectroscopic evidence 
existing at the time which favoured the smaller $q$, finding therefore that \kic\ was an Algol system. Now we have a 
reliable $q$ from extensive spectroscopy it is apposite to revisit the light curve analysis.  The spectroscopic 
$q = 0.09109$ is indeed rather different from either choice considered in Paper\,I.

We modelled the {\it Kepler} short-cadence LC from Quarter 2.3, after removal of the pulsations. The data were 
converted to orbital phase and binned into 230 normal points. The bin sizes were chosen to be 0.002 phase units 
during eclipse and 0.02 outside eclipse, in order to adequately sample the eclipse shapes whilst limiting the 
computation time for individual solutions. This material is identical to that considered in section\,5 in Paper\,I, 
and the scatter of the data is dominated by imperfectly removed pulsations, not by Poisson noise. For the modelling 
process we adopted the 2004 version of the Wilson-Devinney (WD) code \citep{1971ApJ...166..605W,1979ApJ...234.1054W}, 
accessed via the {\sc jktwd} wrapper (Paper\,I) and hereafter called {\sc wd2004}.

\begin{table}  
\tabcolsep 1.6 mm
\caption{\label{tab:wdc} Summary of the control and fixed parameters for the 
{\sc wd2004} solution. For further details see the {\sc wd2004} user guide 
\citep{WilsonVanhamme03}. A and B refer to the primary and secondary stars, respectively.}
\begin{tabular}{llr@{\,$\pm$\,}l} \hline
Parameter                           & {\sc wd2004}          & \mc{Value}            \\
\hline
{\it Control and fixed parameters:} \\
{\sc wd2004} operation mode         & {\sc mode}            & \mc{2}                \\
Treatment of reflection             & {\sc mref}            & \mc{2 (detailed)}     \\
Number of reflections               & {\sc nref}            & \mc{2}                \\
Limb darkening law                  & {\sc ld}              & \mc{3 (sqrt)}         \\
Numerical grid size (normal)        & {\sc n1, n2}          & \mc{60, 50}           \\
Numerical grid size (coarse)        & {\sc n1l, n2l}        & \mc{40, 30}           \\[3pt]
{\it Fixed parameters:} \\
Mass ratio                          & {\sc rm}              & \mc{0.09109}          \\
$T_{\rm eff}$ star\,A (K)           & {\sc tavh}            & \mc{7764}             \\
Rotation rate for star\,A           & {\sc f1}              & \mc{0.756}            \\
Rotation rate for star\,B           & {\sc f2}              & \mc{1.039}            \\
Gravity darkening exponents         & {\sc gr1, gr2}        & \mc{1.0}              \\
Bolometric linear LDC for star\,A   & {\sc xbol1}           & \mc{0.255}            \\
Bolometric nonlinear LDC star\,A    & {\sc ybol1}           & \mc{0.480}            \\
Bolometric linear LDC for star\,B   & {\sc xbol2}           & \mc{0.155}            \\
Bolometric nonlinear LDC star\,B    & {\sc ybol2}           & \mc{0.559}            \\
Passband nonlinear LDC for star\,A  & {\sc y1a}             & \mc{0.062}            \\
Passband linear LDC for star\,B     & {\sc x2a}             & \mc{0.600}            \\
Passband nonlinear LDC for star\,B  & {\sc y2a}             & \mc{0.677}            \\
Third light                         & {\sc el3}             & \mc{0.0}              \\
Orbital eccentricity                & {\sc e}               & \mc{0.0}              \\
Phase shift                         & {\sc pshift}          & \mc{0.0}              \\
\hline\vspace{2mm}
\end{tabular} 
\end{table}

\begin{table}
\tabcolsep 1.5 mm
\caption{\label{tab:wdf} Summary of the fitted and derived parameters.}
\begin{tabular}{llr@{\,$\pm$\,}l} \hline
Parameter                           & {\sc wd2004}          & \mc{Value}            \\
\hline
{\it Fitted parameters:} \\
Star\,A potential                   & {\sc phsv}            &     2.599 & 0.005     \\
Star\,B potential                   & {\sc phsv}            &     1.986 & 0.004     \\
Orbital inclination (\degr)         & {\sc xincl}           &     82.39 & 0.23      \\
$T_{\rm eff}$ star\,B (K)           & {\sc tavc}            &      6001 & 100       \\
Light from star\,A                  & {\sc hlum}            &    0.1354 & 0.0003    \\
Passband linear LDC for star\,A     & {\sc x1a}             &     0.160 & 0.018     \\
Bolometric albedo star\,A           & {\sc alb1}            &      2.46 & 0.32      \\
Bolometric albedo star\,B           & {\sc alb2}            &      0.61 & 0.10      \\
{\it Derived parameters:} \\
Light from star\,B                  & {\sc clum}            &    0.0116 & 0.0008    \\
Fractional radius of star\,A        &                       &    0.4042 & 0.0016    \\
Fractional radius of star\,B        &                       &    0.1764 & 0.0029   \\
{\it Physical properties:} \\
Mass of star\,A (\Msun)               &                     &     2.100 & 0.028     \\
Mass of star\,B (\Msun)               &                     &    0.1913 & 0.0025    \\
Radius of star\,A (\Rsun)             &                     &     2.575 & 0.015     \\
Radius of star\,B (\Rsun)             &                     &     1.124 & 0.019     \\
Surface gravity star\,A (\lg)         &                     &     3.938 & 0.004     \\
Surface gravity star\,B (\lg)         &                     &     3.618 & 0.015     \\
Orbital semimajor axis (\Rsun)        &                     &     6.375 & 0.027     \\
$\log({\rm Luminosity} A / L_{\sun})$ &                     &     1.335 & 0.013     \\
$\log({\rm Luminosity} B / L_{\sun})$ &                     &     0.161 & 0.026     \\
$M_{\rm bol}$ star\,A (mag)           &                     &     1.412 & 0.033     \\
$M_{\rm bol}$ star\,B (mag)           &                     &     4.346 & 0.064     \\
\hline 
\end{tabular} 
\end{table}

 Our immediate conclusion was that fixing $q = 0.09109$ destroys the quality of the fit under the assumption of a 
semi-detached configuration. This mass ratio causes the secondary star to be too large to match the data,  
and the rms of the residuals around the best fit changes from 0.4\,mmag to 10\,mmag. We therefore 
reconsidered the possibility that the system is actually detached, thus removing the requirement for the secondary star 
to fill its Roche lobe. This led to an acceptable fit to the LC, with parameters which are nevertheless very close to 
those found in Paper\,I.

Table\,\ref{tab:wdc} gives a summary of all control and fixed parameters. For our final solutions we used mode 2 in 
{\sc wd2004}, which corresponds to a detached configuration with the light contributions of the two stars governed 
by their radii and \Teff s. We fixed the \Teff\ of star\,A to 7764\,K as determined from spectral analysis and 
included the \Teff\ of star\,B as a fitted parameter. The $K_p$ passband is not implemented in {\sc wd2004}, but 
this does not present a problem as we are studying single-band data. Solutions were calculated using the Johnson 
$V$, $R$ and $I$ and Cousins $R$ and $I$ passbands, allowing us to verify that this mismatch between passbands is 
unimportant. This also helped us to explore the intrinsic precision of the solution by comparing multiple 
alternatives with slightly different starting assumptions. The solutions for the $V$ and $R$ passbands yield 
\Teff s for star\,B in very good agreement with the spectroscopic value found above.

Use of the detailed treatment of reflection in {\sc wd2004} had a negligible effect on the results. We nevertheless 
adopted the detailed treatment as it is physically more realistic. 
The mass ratio was fixed at 0.09109 and a circular orbit was assumed.
 The attempt to include the rotational parameters {\sc f1} and {\sc f2} as fitted parameters led to 
unphysical results. They were fixed to values of 0.756 and 1.039 as found below, where a value of 
1.0 means the star is rotating synchronously with the orbital motion.

Limb darkening was incorporated using the square-root law; the linear limb darkening coefficient (LDC) for star\,A 
was fitted and the other three coefficients were fixed to values obtained from interpolation within the tables of 
\citet{1993AJ....106.2096V}.  This approach was based on preliminary fits which showed that there is 
insufficient information in the LC to fit for the coefficients which we fixed, and on past experience which showed 
that the two coefficients for one star are highly correlated \citep[e.g.][]{Me++07aa}. The other 
fitted parameters were the orbital inclination, the potentials of the two stars, and the passband-specific light 
from star\,A. 

The values of the gravity darkening exponents do not have a significant effect on the  geometric parameters of 
the fit so they were fixed to 1.0. The albedos were included as fitted parameters, and a large value for star\,A 
is required to fit the data. This situation was noted in Paper\,I and implies that {\sc wd2004} is slightly 
struggling to fit data of such high quality as we have here.  The albedos are strongly correlated with the 
gravity darkening exponents; including the gravity darkening exponents as fitted parameters yields lower values 
for the albedos but unreasonable values for the gravity darkening exponents. We interpret this as a limitation of 
the radiative physics included in the {\sc wd2004} model.

\begin{figure} 
\includegraphics[width=9cm]{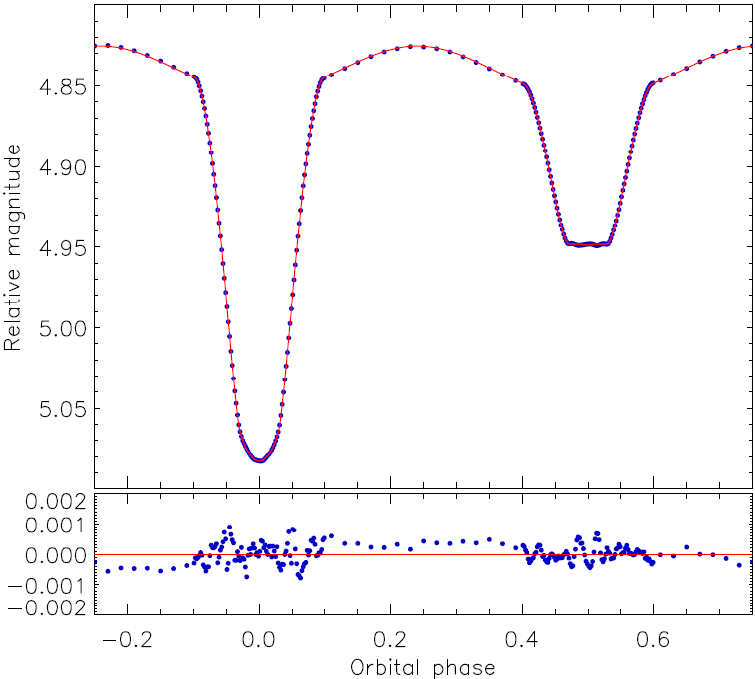}
\caption{\label{fig:wd} Best fit of the phase-binned {\it Kepler} short-cadence
data of \kic\ using {\sc wd2004}. The fitted data are shown in the upper 
part of the plot and the residuals in the lower part of the plot.} 
\label{LC}
\end{figure} 
 
The final results of the modelling process are given in Table\,\ref{tab:wdf}. We have assigned an errorbar to 
each fitted parameter based on the variation of its value between the many acceptable solutions obtained throughout 
the modelling process. These errorbars are in all cases significantly larger than the formal errors returned by 
{\sc wd2004}, which are based on the covariance matrix. Past experience shows that the formal errors are optimistic, 
at least in part because they neglect correlations between parameters 
\citep[Paper I,][]{2009MNRAS.400..791P,2012MNRAS.424L..21T}. 

The vital results from this section are the conclusion that KIC\,10661783 is a detached binary system, and the 
measurement of three photometric parameters from its LC. These three parameters are the orbital inclination and 
the fractional radii of the two stars (i.e.\ the absolute radii divided by the orbital semimajor axis). The best 
fit to the data is shown in Fig.\,\ref{fig:wd}.

The physical properties of the system follow from the spectroscopic and photometric quantities determined above. 
We have calculated the physical properties from the velocity semiamplitudes, effective temperatures and fractional 
radii of the two stars, and the orbital inclination of the system. This was done with the {\sc absdim} code 
\citep{2005A&A...429..645S} and using the set of physical constants tabulated by \cite{2011MNRAS.417.2166S}. 
Uncertainties were propagated within this code by a perturbation approach.

\section{The Roche geometry of the system}

In Paper\,I it was suspected that \kic\ is a semi-detached system and could be an oEA star. Using the newly 
determined mass ratio, we now find that its light curve can be modelled in a satisfactory way only if we adopt a 
detached configuration. We now confirm this finding directly from the Roche model, and exclude signs of activity 
in the system by examining spectra taken at different orbital phases.

\begin{figure}
\includegraphics[angle=-90, width=8.55cm]{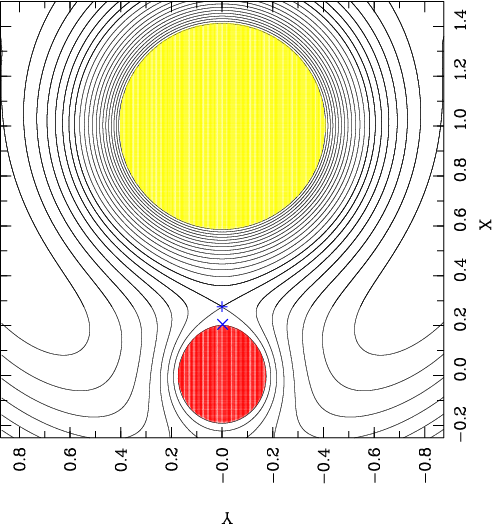}
\caption{Cut through the Roche equipotential surfaces in the x-y plane.
The primary component is shown in yellow, the secondary in red colour. 
The plus sign marks the $L_1$-point, the cross the $P_2$-point.}
\label{Roche_figure}
\end{figure}

In the Roche model and under the assumption of centrally condensed stars and aligned rotational axes, the Roche 
lobe filling factor only depends on the mass ratio and separation of the stars, and on their radii and 
synchronisation factors. We calculated the Roche potential of the binary according to the formulae given in 
Appendix B. From the determined radii and \vs\ we obtain synchronisation factors 
$s = P_{\rm orb}/P_{\rm rot}$ of $s_1 = 0.756 \pm 0.038$ and $s_2 = 1.039 \pm 0.067$ for the primary and 
secondary component, respectively. The rotation of the secondary is consistent with synchronous whereas the 
primary rotates with a significantly subsynchronous velocity. Fig.\,\ref{Roche_figure} shows a cut through the 
equipotential surfaces of the Roche potential in the x-y (orbital) plane for the case of synchronous rotation 
of the secondary. All axes are in units of the separation, and the inner Lagrangian point $L_1$ and the 
substellar surface point of the secondary $P_2$ are marked. It can be seen that the secondary does not fill 
its Roche lobe (RL). The filling factor, defined as the ratio of $P_2$ to $L_1$, is 0.72. Even if we 
vary all input values, i.e.\ $q$, $a$, $R_1$, $R_2$, and $s_1$ within their $1\sigma$ error limits, we get 
a filling factor not larger than 0.84. We therefore reaffirm our conclusion that the system is detached.

\begin{figure}\centering
\includegraphics[width=7cm]{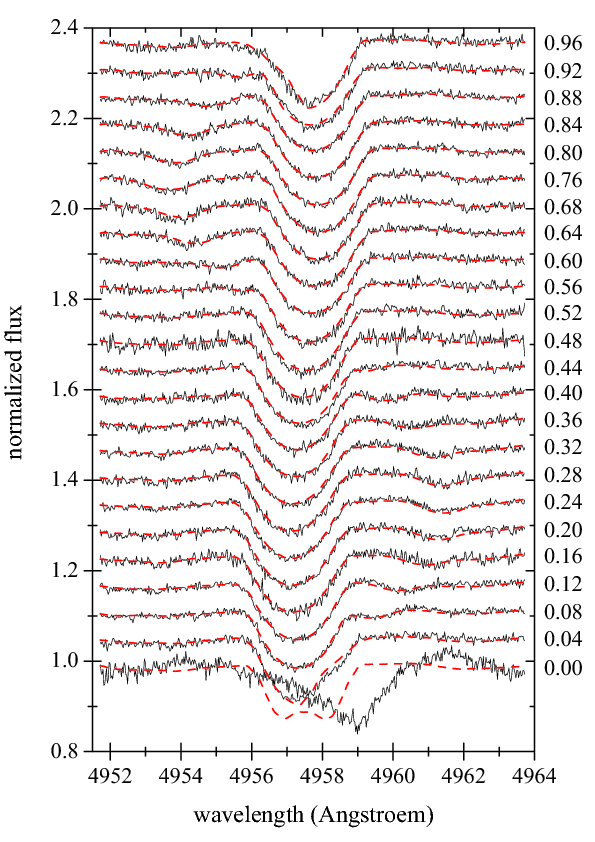}
\caption{Composite profiles from 25 orbital phase bins of the \ion{Fe}{II} 4957\,\AA\ line (black) and
the fit with Shellspec (red). The orbital phase is given to the
right.} 
\label{LPV1}\vspace{3mm}
%
\includegraphics[width=7cm]{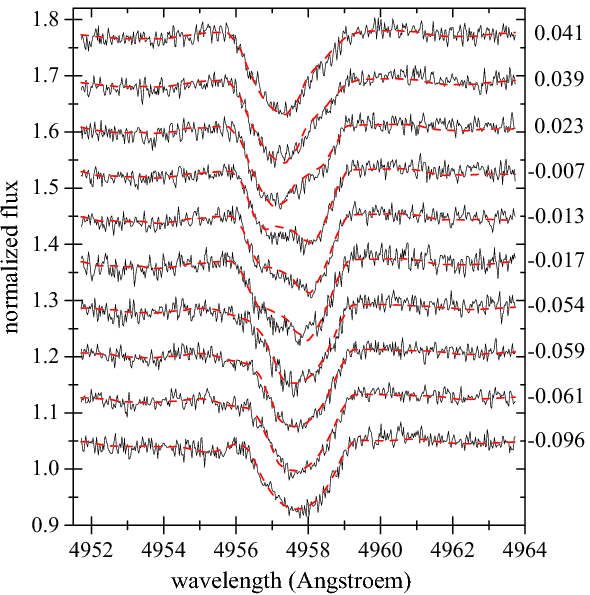}
\caption{As Fig.\,\ref{LPV1} but for single spectra around primary eclipse.} 
\label{LPV2}
\end{figure}

\citet{1989SSRv...50...79K} suggested, from extensive observations with the IUE satellite, that gas flows 
are present in almost all close binaries, even when the evolved star does not seem to fill its critical RL 
\citep[see also][]{1993AJ....106..759O}. To check for possible signs of activity, we did one further test 
using the {\sc Shellspec} program \citep{2004CoSka..34..167B,2005ApJ...623..411B} in its modified version 
{\sc Shellspec07\_inverse} \citep{2009A&A...504..991T}. The program is able to fit the line profiles from 
a time series of composite spectra using two optimised synthetic spectra. It also accounts for the Roche 
geometry of close binaries. We used the program to compute synthetic profiles of the \ion{Fe}{II} 
4957\,\AA\ line based on the parameters listed in Table\,\ref{t6} and the last column of Table\,\ref{t3}.
Then we divided the orbit into 25 phase bins and coadded the spectra whenever more than one spectrum 
falls into the corresponding bin. In the result, each bin represents between one and four spectra.
Fig.\,\ref{LPV1} shows that 
the observed profiles are very well fitted by those computed with {\sc Shellspec}, except for the spectrum 
at the bottom. The reason for the discrepancy very near to primary eclipse is that the line profiles change 
too rapidly in shape at this phase to be averaged over the bin size used. Fig.\,\ref{LPV2} shows the fit of 
the single observed profiles around primary eclipse and it can be seen that also here our model works well. 
The {\sc Shellspec} program allows the O-C residuals to be inspected for mass-transfer effects like gas 
streams or an accretion disc, or remnants of past activity like an inhomogeneous distribution of dense 
circumbinary matter typical for semi-detached (Algol-type) systems. None of these effects could be found 
here: the two-dimensional (wavelength versus orbital phase) O-C distribution 
\citep[see][]{2009A&A...504..991T,2010AJ....139.1327T} is totally flat. Moreover, the almost perfect fit 
of the line profiles during primary eclipse (Fig.\,\ref{LPV2}) confirms our assumption that the rotation 
axis is aligned with the orbital axis (the question of spin-orbit alignment was recently raised due to new 
observations of the detached EB \object{AS Cam}, \citep[see][]{2011ApJ...734L..29P,2009Natur.461..373A}.

\section{Discussion\label{Disc}}

{\it Evolution.} Although we actually observe a detached system it is clear that the secondary component, with a 
mass of 0.2\Msun, a radius of 1.1\Rsun\ and a \te\ of 6000\,K, can only result from mass loss in the past. 
Compared to ``typical'' Algol-type systems, \kic\ has a short orbital period, a very low mass ratio and a primary 
star (the former mass-gainer) which rotates subsynchronous. There is a small group of stars that is characterised 
by the combination of short period, low mass ratio, oversized secondary and overluminous components.  The 
stars have been frequently discussed as the subgroup of the R\,CMa stars, although this subgroup is regarded skeptically
because of the inaccuracies of many of the assigned parameter values at the time it was established
 (see \citealt{1999AJ....117.2980V} for a discussion 
and \citealt{2011MNRAS.418.1764B} for \object{R CMa}). Such binaries are thought to develop according to the so-called 
case B scenario \citep{1969Ap&SS...3...14Z,1971ARA&A...9..183P} where the Roche lobe overflow starts when the 
donor star reaches its critical Roche lobe during rapid expansion after the exhaustion of hydrogen in its centre, i.e. 
prior to core helium burning. The secondary component in this state is oversized and overluminous and its thermonuclear 
power is generated in a hydrogen-burning shell surrounding a degenerate helium core. Possible conservative evolutionary 
scenarios have been discussed by e.g.\ \citet{2001ApJ...552..664N} and \citet{2002ApJ...575..461E}. 
\citet{2008ASPC..392...59S} developed an evolutionary model for \object{OGLEGC 228}, a system that closely resembles many of 
the parameters of the \kic\ system (see Table\,\ref{compstars} and below). The pure conservative model predicts 
progenitors of about equal masses of around 0.9\,\Msun\ in an orbit with only a slightly longer period than actually 
observed.  On the other hand, due to the thick envelope reacting to accretion, such low-mass systems as the R\,CMa stars 
should rapidly evolve into contact configurations 
\citep[see][]{1977ApJ...211..486W,1988AcA....38...89S,1989SSRv...50..205B,2011MNRAS.418.1764B} so that non-conservative 
evolution with mass and angular momentum loss from the system can be assumed \citep{1985ApJ...297..250T}. To avoid the 
strong influence of the convective zone, \citet[][ SDG hereinafter]{1996QJRAS..37...11S}, estimate that the initial mass 
of the secondary must be higher than 2.5\,\Msun\ so its observed remnant mass exceeds 0.3\,\Msun. 
 
{\it Chemical abundance.} The iron abundance as well as the averaged metal abundance of the primary are solar within 
the measurement errors. For the secondary, both values are about 0.3 dex below solar. The carbon abundance determined 
for the primary of \kic\ is $-0.14 \pm 0.28$\,dex below the solar value (i.e.\ solar within the large errorbars); 
that of the secondary component is depleted ($-0.79 \pm 0.25$\,dex relative to solar). Both are typical values for 
Algol systems, as can be seen from a compilation of nine Algols with derived carbon abundances of the primary and 
15 Algols with carbon abundances of the secondary given by SDG. In this compilation, the carbon abundance of the primary 
ranges from $-0.41$ to $+0.06$ dex ($-0.2 \pm 0.2$\,dex on average). For the secondary, carbon is depleted in all cases, 
with an average value of $-0.7\pm 0.4$\,dex. The nitrogen abundance of \kic\ could only be determined for its primary component. 
The enhanced value ($+1.21\pm 0.34$\,dex above solar) is expected by SDG from evolutionary models for the secondary 
component; no estimations for the primary are given.

\begin{table}
\caption{Comparison of object parameters. Values closest to those of KIC = KIC\,10661783 are in bold face.}
\tabcolsep 0.9mm
\begin{tabular}{lcccccccc}
\hline\hline
object        & $P$   &$q$   & $m_1$ & $m_2$ & $R_1$ & $R_2$ &$T_{\rm eff1}$ & $T_{\rm eff2}$\\
              &  (d) & & (\Msun) & (\Msun) & (\Rsun) & (\Rsun) & (K) & (K)\\
	      \hline
KIC           & \bf{1.2314} & \bf{0.093} & \bf{2.05}  & \bf{0.19}  & 2.56  & \bf{1.12}   & \bf{7760}  & \bf{5980}\\
\object{R CMa}        & \bf{1.1359} & 0.170       & 1.07       & \bf{0.17}  & 1.50  & \bf{1.15}   & 7290        & 4250\\
\object{AS Eri}       & 2.6642       & \bf{0.110} & \bf{1.92} & \bf{0.21}  & 1.57  & 2.19         & 8470        & 5110\\
V\,228        & \bf{1.1507} & 0.132       & 1.51       & \bf{0.20}  & 1.36  & \bf{1.24}   & \bf{8070}  & \bf{5810}\\ 
\hline
\end{tabular}
\label{compstars}
\end{table}

{\it Comparison with known Algol-type systems.} \citet{2006MNRAS.373..435I} compiled well-determined absolute parameters 
of Algol-type stars. We searched their compilation for systems with extremely low mass ratios. The stars with the lowest
mass ratios in the list of 74 detached binaries are \object{AR Cas} ($q = 0.315$) and \object{Alpha CrB} ($q = 0.357$). Both stars are 
EBs consisting of two main sequence (MS) stars. Among the 61 semidetached systems we find 14 candidates with $q < 0.2$. 
Restricting the selection to short-period Algols with $P < 5$\,d, nine systems remain. The radius of the secondary of only 
four of these, namely \object{R CMa}, \object{AS Eri}, \object{ST Per}, and \object{XZ Sgr} is less than 3\Rsun. The 
parameters of our object closest to those of \object{R CMa} and \object{AS Eri}, as can be seen from Table\,\ref{compstars}. We added one 
further object, \object{OGLEGC 228} found by \citet{2007AJ....134..541K} as the blue straggler V228 in the globular 
cluster 47\,Tuc. This semi-detached Algol system closely resembles the short period, the low secondary-star mass and radius, 
and the temperatures of both components of \kic, whereas the total mass of \kic\ is very close to that of 
\object{AS Eri}. 
\kic\ differs by one property from all previously found R\,CMa-type objects, however, in that it is a detached system.

\section{Conclusions}

We successfully disentangled the spectra of the primary and the faint secondary component of \kic. The analysis of the 
decomposed spectra gave us the effective temperatures of the stars, and radial velocities derived with {\sc todcor}
yielded an accurate measurement of the mass ratio of components. These quantities were used as input parameters to a new 
modelling of the {\it Kepler} light curve, resulting in the first measurement of the masses and radii of the stars. The 
\te\ of the secondary determined from the LC analysis is $6000 \pm 100$\,K, in good agreement with the spectroscopically 
derived value of $5980 \pm 72$\,K. We also included the determination of the continuum flux ratio between the stars into 
the spectral analysis. The flux ratio of 0.062 at 5000\,\AA\ derived in this way is in agreement within the error bars 
with the $V$-band light ratio of 0.065 calculated from the LC solution.

The main conclusion from the combined spectroscopic-photometric analysis is that the system is detached. The measured 
mass ratio and orbital separation allowed us, together with the finding that the secondary component rotates synchronously, 
to confirm this conclusion directly using the Roche model. 
 The maximum filling factor of the secondary component within the 1$\sigma$ errors of all input parameters
is 0.84. Independently of this finding,
no signs of emission effects, associated with the presence of 
circumbinary matter and often seen in Algol binaries, could be found from applying the {\sc Shellspec} program to the 
observed composite spectra.

The short orbital period and the very low mass ratio make \kic\ a member of the `R\,CMa-type group'. It is, however, the 
only R\,CMa-like system observed so far that is in a detached state and rotates subsynchronously. There are tenable reasons 
for these properties, such as mass and angular momentum loss by non-conservative mass transfer and magnetic braking. 
The unusual combination of short period and low mass ratio in this group may also be related to angular momentum 
redistribution in the presence of third bodies as suspected by e.g.\ \cite{2011MNRAS.418.1764B}. This could be relevant 
to the present case, where orbital angular momentum may well have been redistributed in this old Algol, while tidal 
coupling, given the low mass of the secondary, may have been insufficient to spin up the primary. Other possible 
scenarios have been discussed in Sect.\,\ref{Disc}.
 
A detailed evolutionary model of the system is beyond the scope of this paper. We instead looked for comparable objects 
in the literature. The star that best resembles the parameters of our system is \object{OGLEGC 228}, the blue straggler V228 in 
the globular cluster 47\,Tuc. An attempt by \citet{2008ASPC..392...59S} to derive a model of its evolution was limited 
by the inclusion of pure conservative mass transfer. Models with non-conservative mass transfer, on the other hand, have 
been not applied so far to Algol-type systems of such low mass ratio.

In summary, \kic\ is the short-period Algol-type star with the lowest mass ratio ever observed. Although it has 
properties typical of an R\,CMa object and the primary shows $\delta$\,Sct-type oscillations like the oEA stars, it 
differs from both groups by the fact that it is a detached system. Whereas the stellar parameters of the primary component 
derived in this work will support future asteroseismic modelling of its pulsations, the unusually low mass of the secondary 
will be a challenge for future evolutionary modelling.

\begin{acknowledgements} 
This research is based on observations obtained with the HERMES spectrograph, which is supported by the Fund for Scientific 
Research of Flanders (FWO), Belgium, the Research Council of K.U.Leuven, Belgium, the Fonds National Recherches Scientific 
(FNRS), Belgium, the Royal Observatory of Belgium, the Observatoire de Gen\`eve, Switzerland and the Th\"uringer 
Landessternwarte Tautenburg, Germany. It has made use of the Vienna Atomic Line Database 
\citep[VALD,][]{2000BaltA...9..590K}. JS acknowledges financial support from STFC in the form of an Advanced Fellowship. 
AT acknowledges funding from the European Research Council under the European Community's Seventh Framework Programme 
(FP7/2007-2013)/ERC grant agreement no.\,227224 (PROSPERITY). The authors thank the referee, Prof.
Edwin Budding, for his very useful comments that helped us to improve the article remarkably.
\end{acknowledgements}

\bibliographystyle{aa}          
\bibliography{AA2013_21400.bib} 

\begin{appendix}
\section{Determination of the flux ratios}
\subsection{Basics}
Let $F_k,~k=1,2$ \ be the line fluxes of the two components of the binary and $F_k^c$ the corresponding continuum 
fluxes. The observed composite spectrum normalised to the common continuum of all components is 
\begin{equation}
R = \frac{\d 1}{\d F^c}\d\sum{F_k},~~~F^c = \d\sum{F_k^c}.
\label{a1}
\end{equation}
{\sc korel} delivers the decomposed spectra normalised to the common continuum of the composite input spectrum,
\begin{equation}
R'_1 = \frac{\d F_1+F_2^c}{\d F^c},~~R'_2 = \frac{\d F_2+F_1^c}{\d F^c},~~\sum{R'_k}=R+1.
\label{a2}
\end{equation}
The wavelength-dependent ratios of the single continuum fluxes to the total flux are
\begin{equation}
f_k(\lambda) = \frac{F_k^c}{\d F^c},~~~\d\sum{f_k}=1.
\label{a3}
\end{equation}
We want to calculate the decomposed spectra normalised to the continua of the single components, $R_k = F_k/F_k^c$. 
Introducing the line depths, $r'_k=1-R'_k$ and $r_k=1-R_k$, we simply get
\begin{equation}
r'_k = f_k\,r_k.
\label{a6}
\end{equation}
For the comparison with the synthetic spectra, $S_k$, we replace the $r_k$ in Eq.\,(\ref{a6}) by $s_k=1-S_k$. As a 
measure of the goodness of fit, we use
\begin{equation}
\chi^2 = \sum_\lambda{\sum_{k=1}^2{\frac{\d(r'_k-f_k\,s_k)^2}{\d\sigma^2_k}}}
\label{a7}
\end{equation} 
where the $\sigma^2_k$ stand for the accuracy of the line depths in the decomposed spectra. Since we compare the 
spectra on the scale of the non-normalised decomposed ones, it will be $\sigma_1=\sigma_2$. Setting $f_1=1-f_2$ and 
derivating Eq.\,(\ref{a7}) with respect to $f_2$, we finally get
\begin{equation}
\sum_\lambda{f_2(\lambda)\,(s_1^2+s_2^2)} = \sum_\lambda{s_2 r'_2-s_1\,(r'_1-s_1)}
\label{a8}
\end{equation}
from which we can determine the optimum value of $f_2$ (usually by developing $f_2$ into a polynomial of low degree in 
$\lambda$ and solving the resulting system of linear equations).

\subsection{The optimum, continuum corrected solution}

During the spectrum reduction, the observed composite spectrum is not exactly normalised to the real local continuum 
but to some pseudo-continuum. We introduce correction factors $\alpha_k(\lambda)$ in the sense
\begin{equation}
r'_k(\lambda) = 1-\alpha_k(\lambda)\,R'_k(\lambda)
\label{a9}
\end{equation}
where $R'_k$ are the wrongly normalised decomposed spectra. The $\alpha_k$ can be determined from spline fits 
$C_k^{obs}$ to the uppermost points in the decomposed spectra $R'_k$ and $C_k^{synth}$ to those in the synthetic 
spectra $S_k$. To compute the pseudo-continua on the scale of the non-normalised decomposed spectra, the latter 
spline fit has to be rescaled according to Eq.\,(\ref{a6}) and we get
\begin{equation}
\alpha_k = \frac{\d 1-f_k\,(1-C_k^{synth})}{\d C_k^{obs}}.
\label{a10}
\end{equation}

The occurrence of the so far unknown $f_k$ in Eq.\,(\ref{a10}) does not complicate the calculations because the 
continuum correction can be included into Eq.\,(\ref{a7}) to solve for the optimum values: Inserting the 
continuum-corrected $r'_k$ into Eq.\,(\ref{a7}) and derivating with respect to $f_2$ gives 
\begin{equation}
\begin{array}{l}
\sum_\lambda{f_2(\lambda)\,(B_1^2+B_2^2)} = \sum_\lambda{B_1\,(A_1+B_1)-A_2 B_2}\vspace{2mm}\\
A_k = 1-\frac{\d R'_k}{\d C_k^{obs}}\vspace{1mm}\\
B_k = \left(1-C_k^{synth}\right)\frac{\d R'_k}{\d C_k^{obs}}-s_k.
\end{array}
\label{a11}
\end{equation}
After solving Eq.\,(\ref{a11}) for $f_2$, $\chi^2$ is calculated from Eq.\,(\ref{a7}). 
This can be done on a grid of synthetic spectra 
$s_k$ to find the minimum in $\chi^2$ and the corresponding optimum atmospheric parameters. The renormalised 
decomposed spectra follow from Eqs.\,(\ref{a6}) and (\ref{a9}) to
\begin{equation}
R_k = 1-\frac{\d 1-\alpha_k R'_k}{\d f_k}.
\label{a13}
\end{equation}

\section{The Roche model}

The Roche model is valid for centrally condensed stars where the gravitational potential can be considered as 
the potential of point masses and, in the standard case, for stars that rotate synchronously with the orbit and 
with rotation axes perpendicularly aligned to the orbital plane. For reference, we give here the expression for 
the Roche potential in the reference frame co-rotating with star\,1 for the more general case that star\,1 rotates 
non-synchronously \citep[see e.g.][]{1999ARep...43..229B}:
\begin{equation}\label{Eq1}
\begin{array}{l}
\Phi(x,y,z) = -\frac{\d 1}{\d r_1}-\frac{\d q}{\d r_2}-\vspace{1mm}\\
~~~~-\frac{\d 1+q}{\d 2}\left[(x-x_c)^2+y^2-(1-s^2)(x^2+y^2)\right].
\end{array}
\end{equation}
All coordinates have their origin in the centre of star\,1 and are in units of the separation $a$ between the two 
stars. $x$ points to star\,2, $z$ along the rotation axis of star\,1, and $y$ lies in the orbital plane to span 
a right-handed coordinate system. $q = M_2/M_1$ is the mass ratio, $r_1=\sqrt{x^2+y^2+z^2}$ the distance to the 
centre of star\,1, $r_2=\sqrt{(1-x)^2+y^2+z^2}$ the distance to star\,2, and $x_c = q/(1+q)$. The synchronisation 
factor is $s=\Omega_{\rm rot}/\Omega_{\rm orb}$.

The Roche potential given by Eq.\,\ref{Eq1}, multiplied by $-1$, can be written as
\begin{equation}\label{Eq2}
\begin{array}{l}
\Psi(x,y,z) = \\
~~~=\frac{\d 1}{\d r_1}+q\left(\frac{\d 1}{\d r_2}-x+\frac{\d x_c}{\d 2}\right)+s^2\,(1+q)\frac{\d x^2+y^2}{\d 2}.
\end{array}
\end{equation}
The shapes of the stars follow from the equipotential surfaces passing through the substellar surface points $P$:
\begin{equation}\label{Eq2a}
\begin{array}{l}
\Psi(P,0,0) = \\
~~~=\frac{\d 1}{\d P}+q\left(\frac{\d 1}{\d 1-P}-P+\frac{\d x_c}{\d 2}\right)+s^2\,(1+q)\frac{\d P^2}{\d 2}.
\end{array}
\end{equation}
The position of the inner Lagrangian point $L_1$ is calculated from setting the derivative 
${\rm d}\Psi(x,0,0)/{\rm d}x$ to zero and solving
\begin{equation}
\frac{\d 1}{\d x^2}+q\left[1-\frac{\d 1}{\d (1-x)^2}\right]-s^2\,(1+q)\,x = 0
\end{equation}
numerically.
The radius $r_s$ that describes a sphere of the same volume as the critical Roche lobe (the surface of the Roche 
potential through $L_1$) can be approximated with an accuracy of 1\% \citep{1983ApJ...268..368E} by
\begin{equation}
r_s = \frac{\d 0.49\,q^{2/3}}{\d 0.6\,q^{2/3}+\ln\left(1+q^{1/3}\right)}.
\end{equation}
For a synchronously rotating star, this radius can be used to check if the star fills its critical Roche lobe 
or not. Instead, we used the filling factor $f = P_2/L_1$ as the more accurate criterion, where $P_2$ is the 
substellar surface point of the secondary component (see Fig.\,\ref{Roche_figure}). It is $f=1$ when the star fills 
its critical Roche lobe, i.e.\ when $P_2$ reaches the Lagrangian point $L_1$.
\end{appendix}

\end{document}